\documentclass[dvips]{article}
\usepackage{supertabular,lscape,epsfig}
\usepackage{amssymb}
\usepackage{amsmath}

\DeclareSymbolFont{ppa}{OT1}{ppl}{m}{it}
\DeclareMathSymbol{\vv}{\mathalpha}{ppa}{'166}

\begin{document}

\newcommand{\dd}{\,{\rm d}}
\newcommand{\ie}{{\it i.e.},\,}
\newcommand{\etal}{{\it et al.\ }}
\newcommand{\eg}{{\it e.g.},\,}
\newcommand{\cf}{{\it cf.\ }}
\newcommand{\vs}{{\it vs.\ }}
\newcommand{\zdot}{\makebox[0pt][l]{.}}
\newcommand{\up}[1]{\ifmmode^{\rm #1}\else$^{\rm #1}$\fi}
\newcommand{\dn}[1]{\ifmmode_{\rm #1}\else$_{\rm #1}$\fi}
\newcommand{\upd}{\up{d}}
\newcommand{\uph}{\up{h}}
\newcommand{\upm}{\up{m}}
\newcommand{\ups}{\up{s}}
\newcommand{\arcd}{\ifmmode^{\circ}\else$^{\circ}$\fi}
\newcommand{\arcm}{\ifmmode{'}\else$'$\fi}
\newcommand{\arcs}{\ifmmode{''}\else$''$\fi}
\newcommand{\MS}{{\rm M}\ifmmode_{\odot}\else$_{\odot}$\fi}
\newcommand{\RS}{{\rm R}\ifmmode_{\odot}\else$_{\odot}$\fi}
\newcommand{\LS}{{\rm L}\ifmmode_{\odot}\else$_{\odot}$\fi}

\newcommand{\Abstract}[2]{{\footnotesize\begin{center}ABSTRACT\end{center}
\vspace{1mm}\par#1\par
\noindent
{~}{\it #2}}}

\newcommand{\TabCap}[2]{\begin{center}\parbox[t]{#1}{\begin{center}
  \small {\spaceskip 2pt plus 1pt minus 1pt T a b l e}
  \refstepcounter{table}\thetable \\[2mm]
  \footnotesize #2 \end{center}}\end{center}}

\newcommand{\TabCapp}[2]{\begin{center}\parbox[t]{#1}{\centerline{
  \small {\spaceskip 2pt plus 1pt minus 1pt T a b l e}
  \refstepcounter{table}\thetable}
  \vskip2mm
  \centerline{\footnotesize #2}}
  \vskip3mm
\end{center}}

\newcommand{\TableSep}[2]{\begin{table}[p]\vspace{#1}
\TabCap{#2}\end{table}}

\newcommand{\MakeTableSepp}[4]{\begin{table}[p]\TabCapp{#2}{#3}
  \begin{center} \TableFont \begin{tabular}{#1} #4 
  \end{tabular}\end{center}\end{table}}

\newcommand{\FigCap}[1]{\footnotesize\par\noindent Fig.\  %
  \refstepcounter{figure}\thefigure. #1\par}

\newcommand{\TableFont}{\footnotesize}
\newcommand{\TableFontIt}{\ttit}
\newcommand{\SetTableFont}[1]{\renewcommand{\TableFont}{#1}}

\newcommand{\MakeTable}[4]{\begin{table}[htb]\TabCap{#2}{#3}
  \begin{center} \TableFont \begin{tabular}{#1} #4 
  \end{tabular}\end{center}\end{table}}

\newcommand{\MakeTableSep}[4]{\begin{table}[p]\TabCap{#2}{#3}
  \begin{center} \TableFont \begin{tabular}{#1} #4 
  \end{tabular}\end{center}\end{table}}

\newenvironment{references}%
{
\footnotesize \frenchspacing
\renewcommand{\thesection}{}
\renewcommand{\in}{{\rm in }}
\renewcommand{\AA}{Astron.\ Astrophys.}
\newcommand{\AAS}{Astron.~Astrophys.~Suppl.~Ser.}
\newcommand{\ApJ}{Astrophys.\ J.}
\newcommand{\ApJS}{Astrophys.\ J.~Suppl.~Ser.}
\newcommand{\ApJL}{Astrophys.\ J.~Letters}
\newcommand{\AJ}{Astron.\ J.}
\newcommand{\IBVS}{IBVS}
\newcommand{\PASP}{P.A.S.P.}
\newcommand{\Acta}{Acta Astron.}
\newcommand{\MNRAS}{MNRAS}
\renewcommand{\and}{{\rm and }}
\section{{\rm REFERENCES}}
\sloppy \hyphenpenalty10000
\begin{list}{}{\leftmargin1cm\listparindent-1cm
\itemindent\listparindent\parsep0pt\itemsep0pt}}%
{\end{list}\vspace{2mm}}

\def\TYLDA{~}
\newlength{\DW}
\settowidth{\DW}{0}
\newcommand{\dw}{\hspace{\DW}}

\newcommand{\refitem}[5]{\item[]{#1} #2%
\def\REFARG{#3}\ifx\REFARG\TYLDA\else, {\it#3}\fi
\def\REFARG{#4}\ifx\REFARG\TYLDA\else, {\bf#4}\fi
\def\REFARG{#5}\ifx\REFARG\TYLDA\else, {#5}\fi.}

\newcommand{\Section}[1]{\section{\hskip-6mm.\hskip3mm#1}}
\newcommand{\Subsection}[1]{\subsection{#1}}
\newcommand{\Acknow}[1]{\par\vspace{5mm}{\bf Acknowledgements.} #1}
\pagestyle{myheadings}

\newfont{\bb}{ptmbi8t at 12pt}
\newcommand{\xrule}{\rule{0pt}{2.5ex}}
\newcommand{\xxrule}{\rule[-1.8ex]{0pt}{4.5ex}}
\def\thefootnote{\fnsymbol{footnote}}
\begin{center}
{\Large\bf The Optical Gravitational Lensing Experiment.\\
\vskip6pt
Population II Cepheids in the Galactic Bulge\footnote{Based on observations
obtained with the 1.3~m Warsaw telescope at the Las Campanas Observatory
of the Carnegie Institution of Washington.}}
\vskip.6cm
{\bf ~~M.~~K~u~b~i~a~k$^1$~~and~~A.~~U~d~a~l~s~k~i$^1$ }
\vskip4mm
$^1$Warsaw University Observatory, Al.~Ujazdowskie~4, 00-478~Warszawa, Poland\\
e-mail: (mk,udalski)@astrouw.edu.pl\\
\end{center}

\Abstract{We present $I$-band light curves of 54 Population~II Cepheids 
identified in the OGLE-II catalog of variable objects in the Galactic bulge 
fields. Their periods range from a fraction of a day to several days. Their 
light curves show very close similarity to the light curves of classical 
Cepheids with periods a few times longer. 

We analyze location of the newly identified Population~II Cepheids in the 
color--magnitude diagram. The position of these objects confirms recent 
discovery that the interstellar extinction toward the Galactic bulge might be 
anomalous. The slope of the reddening line obtained from our sample is in very 
good agreement with the earlier one derived with red clump stars and the one 
derived with RR~Lyr stars in the Galactic bulge. 

Our Population~II Cepheids follow the same period--luminosity ($P$--$L$) 
relation indicating similar status of these objects. We compare the $P$--$L$ 
relation of these stars with the relation of Population~II Cepheid detected by 
the OGLE-II survey in the LMC. Deviations from the $P$--$L$ relation of the 
Galactic bulge Cepheids seem to indicate that these objects are located in 
the Galactic bar similarly to red clump stars.}

\Section{Introduction}
Population II variable stars pulsating with periods between about one and 
several days, and occupying the period range between RR~Lyr and W~Vir stars, 
are observed in globular clusters and in the Galactic disk. In the literature 
they are referred to as ``long period RRab stars'', ``short period type~II 
Cepheids'', ``globular cluster Cepheids'' or ``BL~Her'' variable stars. 
Recent review of their physical characteristics was given by Wallerstein 
(2002). 

On the basis of the morphological character of the observed light curves 
Diethelm (1983) distinguished among them three different groups: (1)~long 
period RR~Lyr objects with RR~Lyr-like light curves but periods longer than 
the traditional limit of about 0.7~day; (2)~short period W~Vir stars (CW 
stars) with humps on ascending and descending branches; and (3)~stars with 
humps on ascending branch and much slower decrease of brightness toward 
minimum (BL~Her stars). In the later paper Diethelm (1990) named these groups 
AHB1, AHB2, and AHB3 (Above Horizontal Branch), respectively, on a physical 
ground that all of them are located in the color--magnitude diagram (CMD) by 
about 1~mag above the horizontal branch. The AHB1 stars are thought to belong 
to halo population with a marked deficiency in metal abundance; AHB2 and AHB3 
stars have metallicity close to solar and are considered to be members of 
intermediate (thick disk) population. They are believed to evolve away from 
the Horizontal Branch toward the Asymptotic Branch. 

Absolute magnitude values, metallicity and period--luminosity ($P$--$L$) 
relations for Population~II short period variable stars were discussed by 
Nemec, Nemec and Lutz (1994), Sandage, Diethlem and Tammann (1994) and 
McNamara (1995). Moskalik and Buchler (1993) calculated a sequence of light 
curves for fundamental mode pulsations of Population~II stars with masses 
0.6~\MS and luminosities 125--150~\LS to match the observed light variations 
of BL~Her stars. 

In this paper we present the results of the search for periodic objects with 
periods between about 0.6~days and a few days, where the short period 
Population~II pulsating stars reside, in the OGLE-II photometric database of 
variable objects in the Galactic bulge fields. From the periodogram analysis 
we selected 54 objects with light curve shapes and period values 
characteristic for the AHB-type stars. High quality of the light curves and 
the fact that all these stars belong to the same (Galactic bulge) 
population, may be useful for theoretical modeling of their pulsation. Our 
sample seems to be numerous enough to provide reasonably dense and complete 
period coverage for comparison with other objects of this kind. 

\Section{Observations}
Present work is based on observations obtained in years 1997--1999 during the 
second phase of the Optical Gravitational Lensing Experiment (OGLE-II), which 
is a long term survey of dense stellar fields. The Warsaw 1.3~m telescope 
installed at the Las Campanas Observatory in Chile (operated by the Carnegie 
Institution of Washington), was equipped with a SITE ${2048\times2049}$ first 
generation CCD camera. Observations were conducted in the scanning mode. Each 
${2048\times8192}$ pixel strip corresponded to ${14\arcm\times57\arcm}$ in the 
sky. 49 fields covered in total about 11 square degrees in the direction of 
the Galactic bulge. 

Each field was typically observed between 200 and 300 times in the 
{\it I}-band with the effective exposure time of 87~s, and a dozen times in 
the {\it V}-band with the effective exposure time of 124~s. The more detailed 
description of the instrumental setup of this phase of the OGLE project can 
be found in Udalski, Kubiak and Szyma{\'n}ski (1997). 

Wo{\'z}niak \etal (2002) analyzed the frames collected during the 1997--1999 
seasons with the Difference Image Analysis (DIA) method, as elaborated by 
Alard and Lupton (1998), Alard (2000) and Wo{\'z}niak (2000). The resulting 
catalog contains about 200~000 candidate variable stars. To find the objects 
with periodic light changes we used both classic Fourier analysis and the 
Phase Dispersion Minimization method (Stellingwerf 1978), the latter being 
better suited for the light curve shapes markedly different from sinusoidal. 
All the variable stars were then phased and selected for further analysis 
after inspection of the individual light curves. 

\Section{Selection of the Objects}
Identification of pulsating objects based solely on the period value, shape of 
the light curve and its amplitude can be difficult and ambiguous, in 
particular, for periods of about 1~day where all the AHB objects mentioned 
above overlap. In the present search we were guided by the general appearance 
of the light curves, amplitude of light variations and parameters of the 
Fourier decomposition of the light curves. 

\renewcommand{\arraystretch}{1.05}
\renewcommand{\TableFont}{\scriptsize}
\MakeTableSepp{|@{\hspace{3pt}}l@{\hspace{3pt}}
|@{\hspace{3pt}}r@{\hspace{3pt}}
|@{\hspace{3pt}}r@{\hspace{3pt}}
|@{\hspace{3pt}}c@{\hspace{3pt}}
|@{\hspace{3pt}}c@{\hspace{3pt}}
|@{\hspace{3pt}}r@{\hspace{3pt}}
|@{\hspace{3pt}}r@{\hspace{3pt}}
|@{\hspace{3pt}}c@{\hspace{3pt}}
|@{\hspace{3pt}}c@{\hspace{3pt}}
|@{\hspace{3pt}}c@{\hspace{3pt}}
|@{\hspace{3pt}}c@{\hspace{3pt}}|}{12cm}{Population II Cepheids
in the Galactic bulge}{\hline
Scan & Cat &OGLE II &R.A.   & DEC   &$l$~~~~ & $b$~~~~ &$V-I$&$I$& Ampl.&Period\\ 
~~\# & \#~~& \#~~~  &(2000.)&(2000.)&    &     & mag &mag& mag  & days \\
\hline
bul\_sc5  & 6900 & 429663 & 17\uph50\upm54\zdot\ups00 &$-29\arcd31\arcm41\zdot\arcs8$&    0.08 & $-1.29$ &  3.346 & 17.683 & 0.42 &  0.66782\\
bul\_sc24 & 1609 & 517291 & 17\uph53\upm44\zdot\ups41 &$-32\arcd57\arcm14\zdot\arcs0$&  357.44 & $-3.55$ &  1.700 & 15.991 & 0.50 &  0.76522\\
bul\_sc17 & 2499 & 104441 & 18\uph10\upm34\zdot\ups17 &$-26\arcd08\arcm39\zdot\arcs3$&    5.18 & $-3.39$ &   --   & 16.806 & 0.42 &  0.83163\\
bul\_sc39 & 2239 & 664382 & 17\uph56\upm05\zdot\ups04 &$-29\arcd54\arcm51\zdot\arcs8$&    0.32 & $-2.45$ &  2.140 & 15.978 & 0.23 &  0.92213\\
bul\_sc22 & 3993 & 332505 & 17\uph56\upm33\zdot\ups03 &$-30\arcd36\arcm33\zdot\arcs5$&  359.77 & $-2.89$ &  2.188 & 15.937 & 0.29 &  0.93110\\
bul\_sc31 &  662 &  23556 & 18\uph01\upm56\zdot\ups92 &$-28\arcd55\arcm11\zdot\arcs6$&    1.82 & $-3.07$ &  1.275 & 15.405 & 0.23 &  0.93951\\
bul\_sc37 & 5099 & 264353 & 17\uph52\upm26\zdot\ups01 &$-29\arcd48\arcm56\zdot\arcs4$&    0.01 & $-1.72$ &  2.484 & 15.483 & 0.19 &  0.95258\\
bul\_sc28 &  523 & 241866 & 17\uph47\upm23\zdot\ups13 &$-37\arcd12\arcm36\zdot\arcs1$&  353.10 & $-4.60$ &  1.295 & 15.092 & 0.31 &  0.97318\\
bul\_sc43 &  351 & 130103 & 17\uph35\upm08\zdot\ups13 &$-27\arcd31\arcm25\zdot\arcs8$&  359.97 & $ 2.70$ &  2.210 & 16.229 & 0.59 &  1.09776\\
bul\_sc14 & 1768 & 233342 & 17\uph46\upm56\zdot\ups84 &$-23\arcd10\arcm04\zdot\arcs4$&    5.08 & $ 2.74$ &  1.871 & 14.891 & 0.57 &  1.20950\\
bul\_sc14 & 1241 & 366395 & 17\uph47\upm16\zdot\ups21 &$-23\arcd16\arcm30\zdot\arcs5$&    5.03 & $ 2.62$ &  1.752 & 14.842 & 0.51 &  1.28364\\
bul\_sc30 & 1107 & 604452 & 18\uph01\upm47\zdot\ups33 &$-29\arcd07\arcm39\zdot\arcs1$&    1.62 & $-3.14$ &  1.333 & 14.286 & 0.46 &  1.33916\\
bul\_sc39 & 3620 & 515806 & 17\uph55\upm43\zdot\ups80 &$-29\arcd44\arcm49\zdot\arcs9$&    0.43 & $-2.30$ &  1.608 & 15.006 & 0.51 &  1.35410\\
bul\_sc47 &  461 & 254614 & 17\uph27\upm24\zdot\ups11 &$-39\arcd51\arcm30\zdot\arcs9$&  348.77 & $-2.71$ &  1.809 & 14.524 & 0.37 &  1.47710\\
bul\_sc3  &  791 & 227653 & 17\uph53\upm27\zdot\ups82 &$-30\arcd19\arcm55\zdot\arcs3$&  359.67 & $-2.17$ &  1.491 & 16.130 & 0.26 &  1.48412\\
bul\_sc5  & 3719 & 146244 & 17\uph50\upm21\zdot\ups49 &$-29\arcd54\arcm27\zdot\arcs5$&  359.70 & $-1.38$ &  3.330 & 16.399 & 0.73 &  1.50105\\
bul\_sc32 & 2167 &  87478 & 18\uph02\upm56\zdot\ups05 &$-28\arcd40\arcm39\zdot\arcs2$&    2.14 & $-3.14$ &   --   & 14.835 & 0.61 &  1.50523\\
bul\_sc4  &  170 & 406186 & 17\uph54\upm38\zdot\ups21 &$-30\arcd10\arcm41\zdot\arcs8$&  359.93 & $-2.32$ &  1.633 & 14.975 & 0.85 &  1.53185\\
bul\_sc44 & 5324 & 200215 & 17\uph49\upm30\zdot\ups95 &$-29\arcd50\arcm58\zdot\arcs2$&  359.65 & $-1.19$ &   --   & 16.884 & 0.56 &  1.55140\\
bul\_sc43 & 2183 & 311829 & 17\uph35\upm13\zdot\ups82 &$-27\arcd03\arcm16\zdot\arcs6$&    0.37 & $ 2.94$ &  2.284 & 15.982 & 0.65 &  1.61993\\
bul\_sc43 & 2694 & 326907 & 17\uph35\upm29\zdot\ups17 &$-26\arcd55\arcm52\zdot\arcs8$&    0.51 & $ 2.96$ &  2.216 & 15.777 & 0.54 &  1.66050\\
bul\_sc21 & 3035 &  91346 & 17\uph59\upm53\zdot\ups52 &$-28\arcd56\arcm37\zdot\arcs5$&    1.58 & $-2.69$ &  1.588 & 14.863 & 0.55 &  1.73020\\
bul\_sc45 & 1189 & 256269 & 18\uph03\upm33\zdot\ups61 &$-30\arcd01\arcm14\zdot\arcs6$&    1.03 & $-3.91$ &  1.189 & 14.457 & 0.77 &  1.74795\\
bul\_sc5  &  400 & 102694 & 17\uph50\upm13\zdot\ups13 &$-30\arcd20\arcm56\zdot\arcs5$&  359.30 & $-1.58$ &  3.373 & 16.754 & 0.40 &  1.78292\\
bul\_sc4  & 8691 & 765688 & 17\uph55\upm06\zdot\ups75 &$-29\arcd18\arcm07\zdot\arcs5$&    0.74 & $-1.96$ &  2.105 & 15.489 & 0.52 &  1.81738\\
bul\_sc23 &  611 & 577789 & 17\uph58\upm14\zdot\ups58 &$-31\arcd33\arcm23\zdot\arcs7$&  359.13 & $-3.68$ &  1.496 & 14.647 & 0.71 &  1.85470\\
bul\_sc43 & 1415 & 159919 & 17\uph35\upm02\zdot\ups07 &$-27\arcd14\arcm50\zdot\arcs6$&    0.19 & $ 2.87$ &  2.548 & 15.628 & 0.38 &  1.91817\\
bul\_sc37 & 3445 &  73374 & 17\uph52\upm15\zdot\ups46 &$-30\arcd00\arcm17\zdot\arcs0$&  359.82 & $-1.78$ &  2.813 & 16.200 & 0.42 &  2.08811\\
bul\_sc4  & 5359 & 120543 & 17\uph54\upm09\zdot\ups10 &$-29\arcd39\arcm58\zdot\arcs7$&    0.32 & $-1.97$ &  1.879 & 14.510 & 0.53 &  2.21270\\
bul\_sc3  & 5274 & 341515 & 17\uph53\upm29\zdot\ups21 &$-29\arcd48\arcm52\zdot\arcs5$&    0.12 & $-1.92$ &  2.006 & 14.887 & 0.60 &  2.21828\\
bul\_sc30 & 4393 & 126172 & 18\uph01\upm04\zdot\ups52 &$-28\arcd41\arcm21\zdot\arcs6$&    1.93 & $-2.79$ &  1.493 & 14.509 & 0.44 &  2.24745\\
bul\_sc37 & 3808 &  81466 & 17\uph52\upm07\zdot\ups85 &$-29\arcd57\arcm01\zdot\arcs0$&  359.86 & $-1.73$ &  3.524 & 16.339 & 0.41 &  2.29300\\
bul\_sc37 & 2246 & 369978 & 17\uph52\upm36\zdot\ups22 &$-30\arcd08\arcm33\zdot\arcs2$&  359.74 & $-1.92$ &  2.257 & 15.107 & 0.43 &  2.30150\\
bul\_sc34 & 4507 & 388943 & 17\uph58\upm03\zdot\ups44 &$-29\arcd01\arcm13\zdot\arcs4$&    1.31 & $-2.38$ &  1.930 & 14.734 & 0.47 &  2.31407\\
bul\_sc34 & 4631 & 159129 & 17\uph58\upm01\zdot\ups53 &$-28\arcd59\arcm56\zdot\arcs6$&    1.33 & $-2.36$ &  1.938 & 15.340 & 0.63 &  2.33010\\
bul\_sc25 &  598 &  26722 & 17\uph54\upm00\zdot\ups19 &$-33\arcd08\arcm22\zdot\arcs9$&  357.30 & $-3.69$ &  2.081 & 15.464 & 0.26 &  2.73418\\
bul\_sc20 &  961 &  29057 & 17\uph58\upm55\zdot\ups93 &$-29\arcd10\arcm57\zdot\arcs0$&    1.26 & $-2.62$ &  1.302 & 14.289 & 0.73 &  2.88445\\
bul\_sc5  & 2688 & 364862 & 17\uph50\upm38\zdot\ups53 &$-30\arcd03\arcm48\zdot\arcs5$&  359.59 & $-1.51$ &  0.842 & 13.496 & 0.17 &  2.99040\\
bul\_sc14 & 3347 & 285762 & 17\uph47\upm01\zdot\ups88 &$-22\arcd51\arcm34\zdot\arcs2$&    5.36 & $ 2.88$ &  1.793 & 14.312 & 0.51 &  3.08870\\
bul\_sc41 & 3841 & 145401 & 17\uph51\upm42\zdot\ups51 &$-32\arcd41\arcm41\zdot\arcs3$&  357.45 & $-3.05$ &  2.150 & 14.624 & 0.63 &  3.20082\\
bul\_sc4  & 8846 & 194227 & 17\uph54\upm06\zdot\ups61 &$-29\arcd16\arcm22\zdot\arcs1$&    0.66 & $-1.76$ &  2.482 & 14.852 & 0.59 &  3.51336\\
bul\_sc4  & 2323 & 635511 & 17\uph54\upm55\zdot\ups52 &$-29\arcd57\arcm31\zdot\arcs0$&    0.16 & $-2.26$ &  1.866 & 15.067 & 0.49 &  3.54254\\
bul\_sc39 &  616 &  14171 & 17\uph55\upm12\zdot\ups35 &$-30\arcd07\arcm24\zdot\arcs1$&    0.04 & $-2.39$ &  1.830 & 14.448 & 0.23 &  3.65017\\
bul\_sc22 &  815 & 220104 & 17\uph56\upm46\zdot\ups47 &$-31\arcd07\arcm07\zdot\arcs3$&  359.35 & $-3.19$ &  2.118 & 15.016 & 0.24 &  4.49944\\
bul\_sc3  & 1755 & 455680 & 17\uph53\upm34\zdot\ups75 &$-30\arcd12\arcm39\zdot\arcs7$&  359.79 & $-2.13$ &  2.023 & 14.811 & 0.31 &  4.57875\\
bul\_sc38 & 3260 & 479529 & 18\uph01\upm32\zdot\ups68 &$-29\arcd49\arcm11\zdot\arcs5$&    0.99 & $-3.43$ &  1.611 & 14.001 & 0.20 &  5.53097\\
bul\_sc16 & 3512 & 648491 & 18\uph10\upm31\zdot\ups33 &$-26\arcd04\arcm56\zdot\arcs7$&    5.23 & $-3.35$ &  2.158 & 15.443 & 0.16 &  5.84966\\
bul\_sc20 & 3867 & 131735 & 17\uph58\upm55\zdot\ups95 &$-28\arcd43\arcm19\zdot\arcs5$&    1.66 & $-2.40$ &  1.683 & 13.814 & 0.29 &  7.13700\\
bul\_sc2  & 1657 & 278722 & 18\uph04\upm19\zdot\ups87 &$-29\arcd02\arcm44\zdot\arcs2$&    1.97 & $-3.58$ &  1.628 & 13.398 & 0.14 &  7.45600\\
bul\_sc25 &  693 & 195800 & 17\uph54\upm24\zdot\ups38 &$-33\arcd06\arcm51\zdot\arcs0$&  357.37 & $-3.75$ &  2.125 & 14.188 & 0.30 &  7.73934\\
bul\_sc40 & 2524 & 258103 & 17\uph50\upm58\zdot\ups92 &$-33\arcd08\arcm53\zdot\arcs1$&  356.98 & $-3.15$ &  2.134 & 14.578 & 0.23 &  7.77000\\
bul\_sc26 &  894 & 584679 & 17\uph47\upm48\zdot\ups40 &$-35\arcd16\arcm06\zdot\arcs7$&  354.82 & $-3.67$ &  1.475 & 13.588 & 0.61 &  8.79121\\
bul\_sc7  &  540 &  42330 & 18\uph08\upm44\zdot\ups36 &$-32\arcd13\arcm10\zdot\arcs6$&  359.63 & $-5.94$ &  1.403 & 14.032 & 0.54 &  9.52109\\
bul\_sc39 & 5385 & 361403 & 17\uph55\upm23\zdot\ups12 &$-29\arcd31\arcm35\zdot\arcs3$&    0.58 & $-2.13$ &  1.880 & 13.309 & 0.77 &  9.94431\\
\hline}
In Table~1 we list all the objects found in the catalog in the period range 
in question and with light curve shapes resembling those described by Diethelm 
(1983). The consecutive columns of Table~1 give: number of the scanned field, 
number in Wo{\'z}niak \etal (2002) catalog, number in OGLE-II database, 
equatorial coordinates (J2000.0), Galactic coordinates ($l,b$), 
mean values of ${(V-I)}$ and $I$, amplitude of light variations, and period in 
days. The mean {\it VI} photometry comes from the OGLE maps of the Galactic 
bulge (Udalski \etal 2002). 

\begin{figure}[htb]
\centerline{\hglue-5mm\includegraphics[bb=60 190 565 690, width=12.5cm]{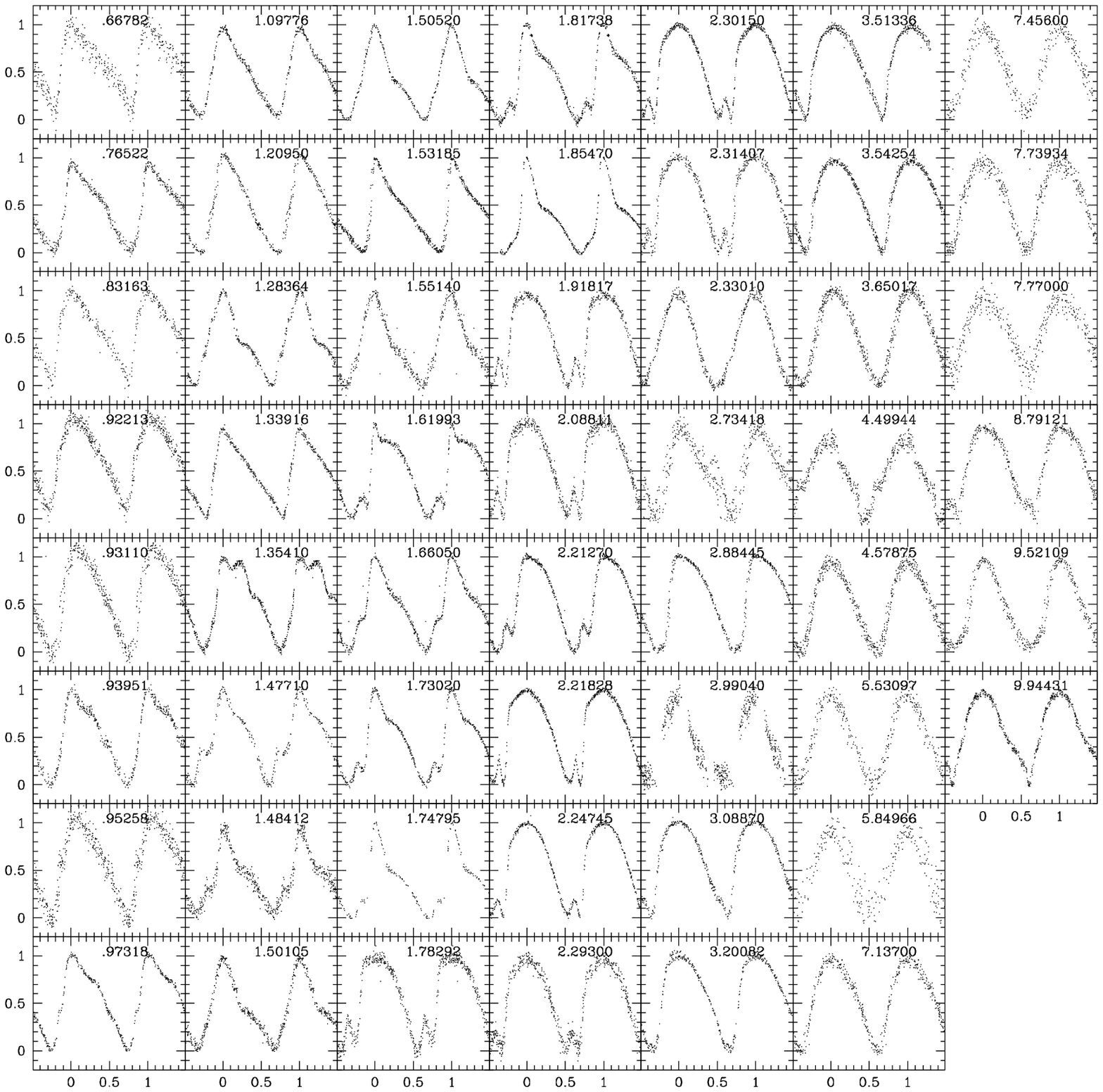}} 
\FigCap{Light curves of the Galactic bulge Population~II Cepheids listed in 
Table~1. Amplitudes are normalized to 1 mag and maxima of brightness  are set at 
phase zero.} 
\end{figure}
Light curves of the stars from Table~1 are shown in Fig.~1. For easier 
comparison of the shapes of particular light curves we normalized them to one 
magnitude amplitude and set the maximum brightness at phase zero. The curves 
are ordered according to the increasing period values (from up to down). 

It is worth noticing that in spite of rather poor quality of formerly 
published photometric observations of the short period Population~II Cepheids 
we may identify in Fig.~1 the light curves with shapes considered to be 
characteristic for all the AHB groups mentioned above. 

\begin{figure}[p]
\vglue-7mm
\centerline{\hglue-5mm\includegraphics[bb=60 190 565 690, width=12.5cm, 
height=19.5cm]{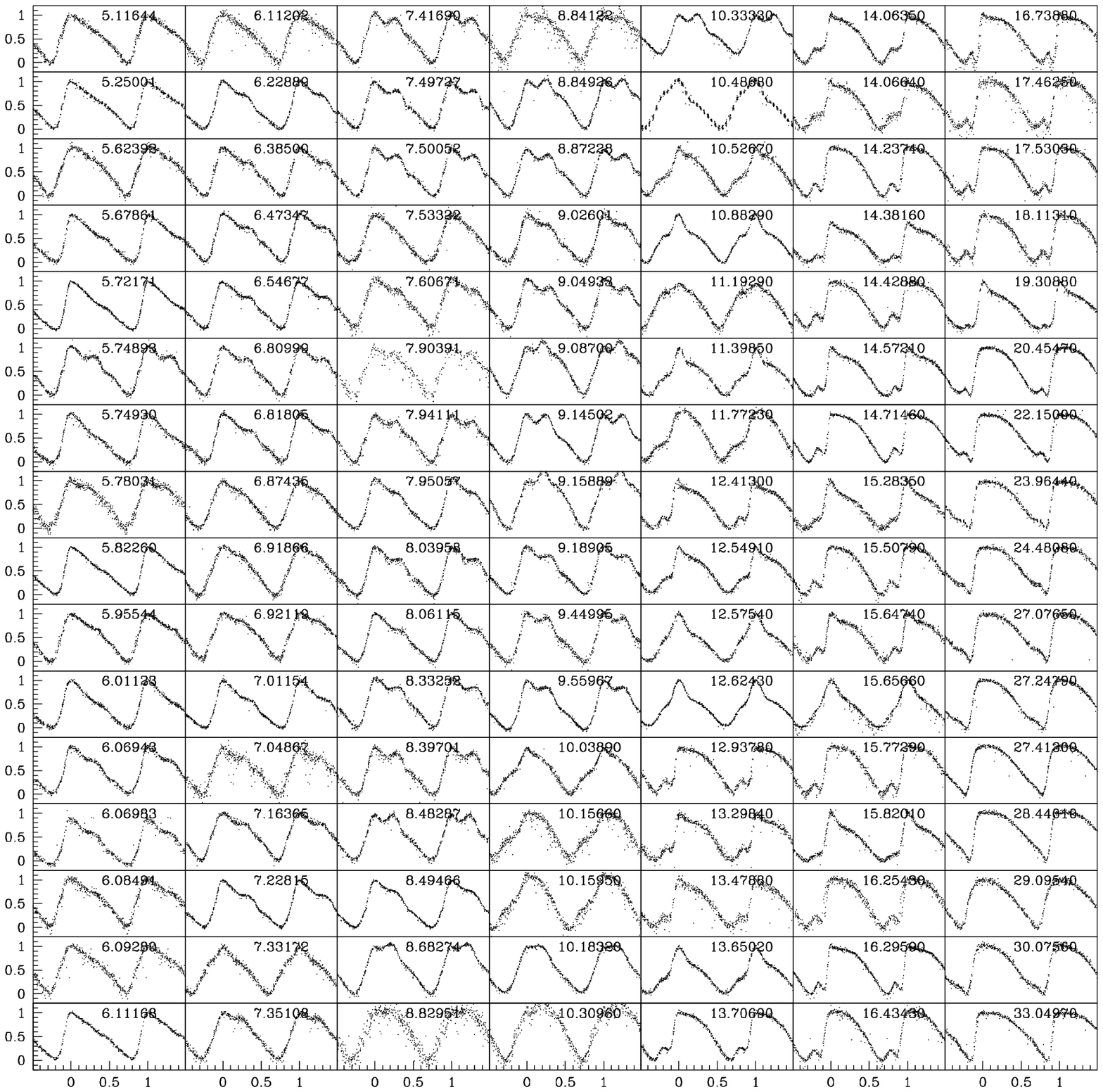}} 
\FigCap{Light curves of the Small Magellanic Cloud classical Cepheids. 
Amplitudes are normalized to 1 mag and maxima of brightness are set at phase 
zero.} 
\end{figure}
\renewcommand{\arraystretch}{.9}
\MakeTable{|l|r|c|c|c|c|c|c|}{12cm}{
Population II Cepheids in the Large Magellanic Cloud}
{\hline
Scan &OGLE II &R.A.&       DEC &            $(V - I)_0$ &  $I_0$&     
Ampl.
& Period\\
\#   &   \#    & (2000.) &  (2000.)&          mag &  mag&   
mag& days \\
\hline
lmc\_sc8  & 82008 &5\uph15\upm08\zdot\ups72 &$-68\arcd54\arcm53\zdot\arcs6$&  0.318 & 17.649& 0.71&  1.15216\\
lmc\_sc7  &127752 &5\uph18\upm35\zdot\ups83 &$-69\arcd45\arcm45\zdot\arcs8$&  0.331 & 17.753& 0.57&  1.16730\\
lmc\_sc10 &256258 &5\uph12\upm30\zdot\ups52 &$-69\arcd07\arcm16\zdot\arcs4$&  0.389 & 17.734& 0.50&  1.18152\\
lmc\_sc9  & 52800 &5\uph12\upm30\zdot\ups52 &$-69\arcd07\arcm16\zdot\arcs3$&  0.373 & 17.730& 0.54&  1.18153\\
lmc\_sc8  &194670 &5\uph16\upm21\zdot\ups55 &$-69\arcd36\arcm59\zdot\arcs4$&  0.121 & 17.520& 0.37&  1.21380\\
lmc\_sc6  &254530 &5\uph21\upm19\zdot\ups79 &$-69\arcd56\arcm56\zdot\arcs3$&  0.354 & 17.499& 0.72&  1.26601\\
lmc\_sc17 &224169 &5\uph40\upm03\zdot\ups18 &$-70\arcd04\arcm47\zdot\arcs8$&  0.312 & 16.733& 0.92&  1.32155\\
lmc\_sc21 &179265 &5\uph21\upm58\zdot\ups43 &$-70\arcd16\arcm35\zdot\arcs2$&  0.410 & 17.363& 0.73&  1.48928\\
lmc\_sc17 &231021 &5\uph39\upm41\zdot\ups14 &$-69\arcd58\arcm01\zdot\arcs4$&  0.470 & 17.391& 0.37&  1.55472\\
lmc\_sc9  &264013 &5\uph14\upm27\zdot\ups14 &$-68\arcd58\arcm02\zdot\arcs2$&  0.422 & 17.418& 0.50&  1.60928\\
lmc\_sc8  &306227 &5\uph17\upm07\zdot\ups60 &$-69\arcd27\arcm34\zdot\arcs1$&  0.525 & 17.504& 0.54&  1.77079\\
lmc\_sc5  &323141 &5\uph23\upm56\zdot\ups01 &$-69\arcd25\arcm30\zdot\arcs1$&  0.677 & 17.554& 0.58&  1.96664\\
lmc\_sc12 & 37723 &5\uph05\upm15\zdot\ups30 &$-69\arcd22\arcm08\zdot\arcs9$&  0.228 & 16.675& 0.36&  2.47593\\
lmc\_sc2  &310497 &5\uph31\upm52\zdot\ups39 &$-69\arcd30\arcm26\zdot\arcs3$&  0.600 & 17.216& 0.62&  2.67170\\
lmc\_sc11 &263536 &5\uph09\upm21\zdot\ups99 &$-69\arcd36\arcm03\zdot\arcs2$&  0.415 & 16.830& 0.42&  3.23438\\
lmc\_sc13 & 30676 &5\uph05\upm10\zdot\ups89 &$-68\arcd47\arcm47\zdot\arcs4$&  0.162 & 15.943& 0.51&  4.01406\\
lmc\_sc3  &290130 &5\uph29\upm28\zdot\ups64 &$-69\arcd48\arcm00\zdot\arcs5$&  0.440 & 16.059& 0.66&  4.07487\\
lmc\_sc17 &237748 &5\uph39\upm50\zdot\ups08 &$-69\arcd50\arcm53\zdot\arcs0$&  1.668 & 17.939& 0.70&  4.14750\\
lmc\_sc12 &100127 &5\uph05\upm44\zdot\ups90 &$-69\arcd14\arcm56\zdot\arcs2$&  0.330 & 16.086& 0.28&  4.92309\\
lmc\_sc14 &170192 &5\uph03\upm49\zdot\ups01 &$-68\arcd55\arcm05\zdot\arcs1$&  0.665 & 16.434& 0.15&  6.37007\\
lmc\_sc8  &218916 &5\uph16\upm29\zdot\ups21 &$-69\arcd24\arcm09\zdot\arcs2$&  0.844 & 15.937& 0.24&  6.71586\\
lmc\_sc19 & 74281 &5\uph43\upm37\zdot\ups52 &$-70\arcd38\arcm05\zdot\arcs2$&  0.597 & 15.884& 0.24&  7.21116\\
lmc\_sc14 &174795 &5\uph03\upm59\zdot\ups25 &$-68\arcd53\arcm24\zdot\arcs1$&  0.523 & 15.507& 0.30&  9.39813\\
lmc\_sc14 &139172 &5\uph04\upm22\zdot\ups45 &$-69\arcd20\arcm42\zdot\arcs6$&  0.760 & 16.022& 0.27&  9.86728\\
lmc\_sc3  &266664 &5\uph29\upm08\zdot\ups36 &$-69\arcd56\arcm04\zdot\arcs3$&  0.578 & 15.606& 0.27& 10.02368\\
lmc\_sc6  &118148 &5\uph20\upm20\zdot\ups71 &$-69\arcd12\arcm21\zdot\arcs0$&  0.734 & 15.943& 0.36& 10.51146\\
lmc\_sc17 &221134 &5\uph39\upm44\zdot\ups56 &$-70\arcd08\arcm21\zdot\arcs9$&  0.931 & 16.016& 0.39& 11.22865\\
lmc\_sc6  &350603 &5\uph21\upm18\zdot\ups99 &$-69\arcd11\arcm47\zdot\arcs5$&  0.631 & 15.816& 0.70& 11.40997\\
lmc\_sc18 &185847 &5\uph42\upm19\zdot\ups16 &$-70\arcd24\arcm08\zdot\arcs2$&  0.880 & 15.965& 0.37& 12.20018\\
lmc\_sc3  &108113 &5\uph27\upm59\zdot\ups92 &$-69\arcd23\arcm27\zdot\arcs6$&  0.798 & 15.913& 0.47& 12.72120\\
lmc\_sc21 &132285 &5\uph21\upm35\zdot\ups39 &$-70\arcd13\arcm25\zdot\arcs8$&  0.829 & 15.769& 0.56& 12.90450\\
lmc\_sc14 &200768 &5\uph04\upm51\zdot\ups93 &$-69\arcd23\arcm55\zdot\arcs9$&  0.767 & 15.622& 0.77& 14.85800\\
lmc\_sc14 &156721 &5\uph04\upm07\zdot\ups79 &$-69\arcd07\arcm31\zdot\arcs8$&  0.936 & 15.989& 0.83& 14.89065\\
lmc\_sc7  &134411 &5\uph18\upm17\zdot\ups91 &$-69\arcd43\arcm27\zdot\arcs8$&  0.673 & 15.435& 0.96& 15.84421\\
lmc\_sc7  &239698 &5\uph19\upm26\zdot\ups55 &$-69\arcd51\arcm51\zdot\arcs1$&  0.508 & 14.870& 0.76& 17.52599\\
\hline}
\begin{figure}[htb]
\vglue-.7cm
\centerline{\includegraphics[bb=60 190 430 690, width=9.3cm]{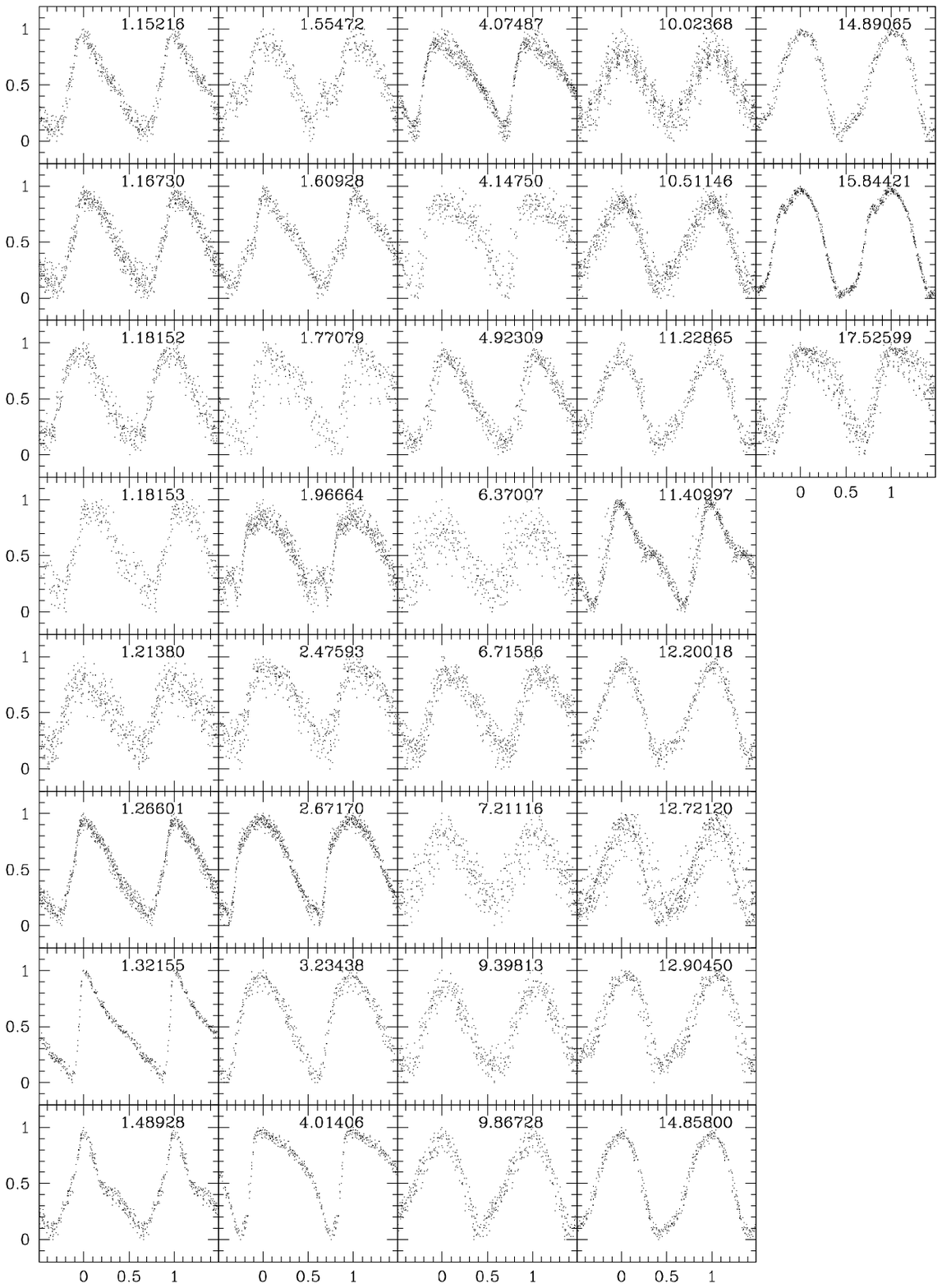}}
\FigCap{Light curves of Population~II Cepheids from the Large Magellanic 
Clouds with amplitudes normalized to 1~mag and maxima of brightness set at phase 
zero.}
\end{figure}
\begin{figure}[p]
\vglue-1.2cm
\centerline{\includegraphics[width=10.9cm]{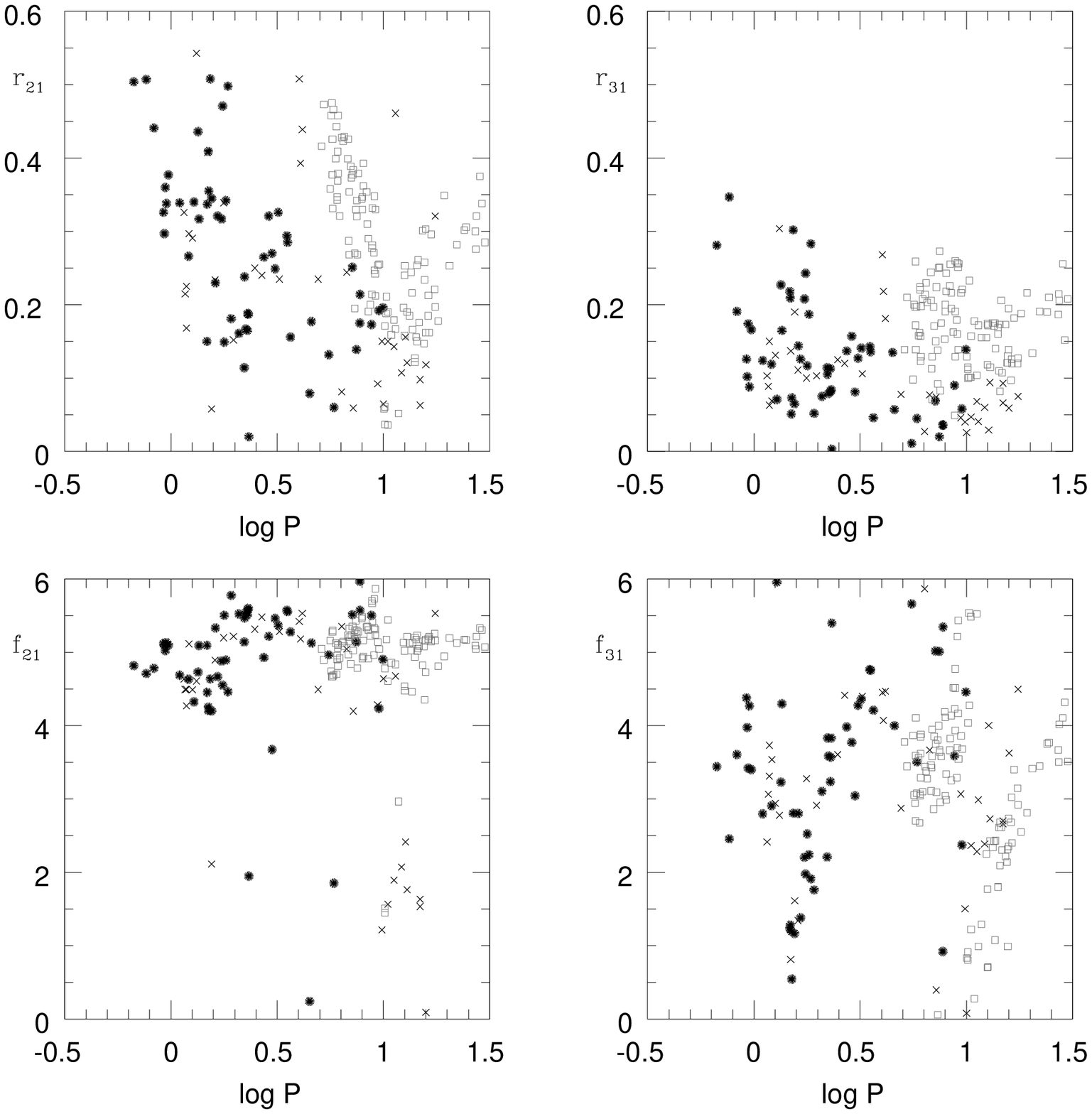}} 
\FigCap{Fourier parameters of the light curves of the objects discussed in the 
paper. Big dots denote the Galactic bulge Population~II Cepheids, crosses -- 
Population~II Cepheids from the Large Magellanic Cloud, and squares -- 
classical Cepheids from the Small Magellanic Cloud, shown in Fig.~2.} 
\vskip7mm
\centerline{\includegraphics[width=6cm]{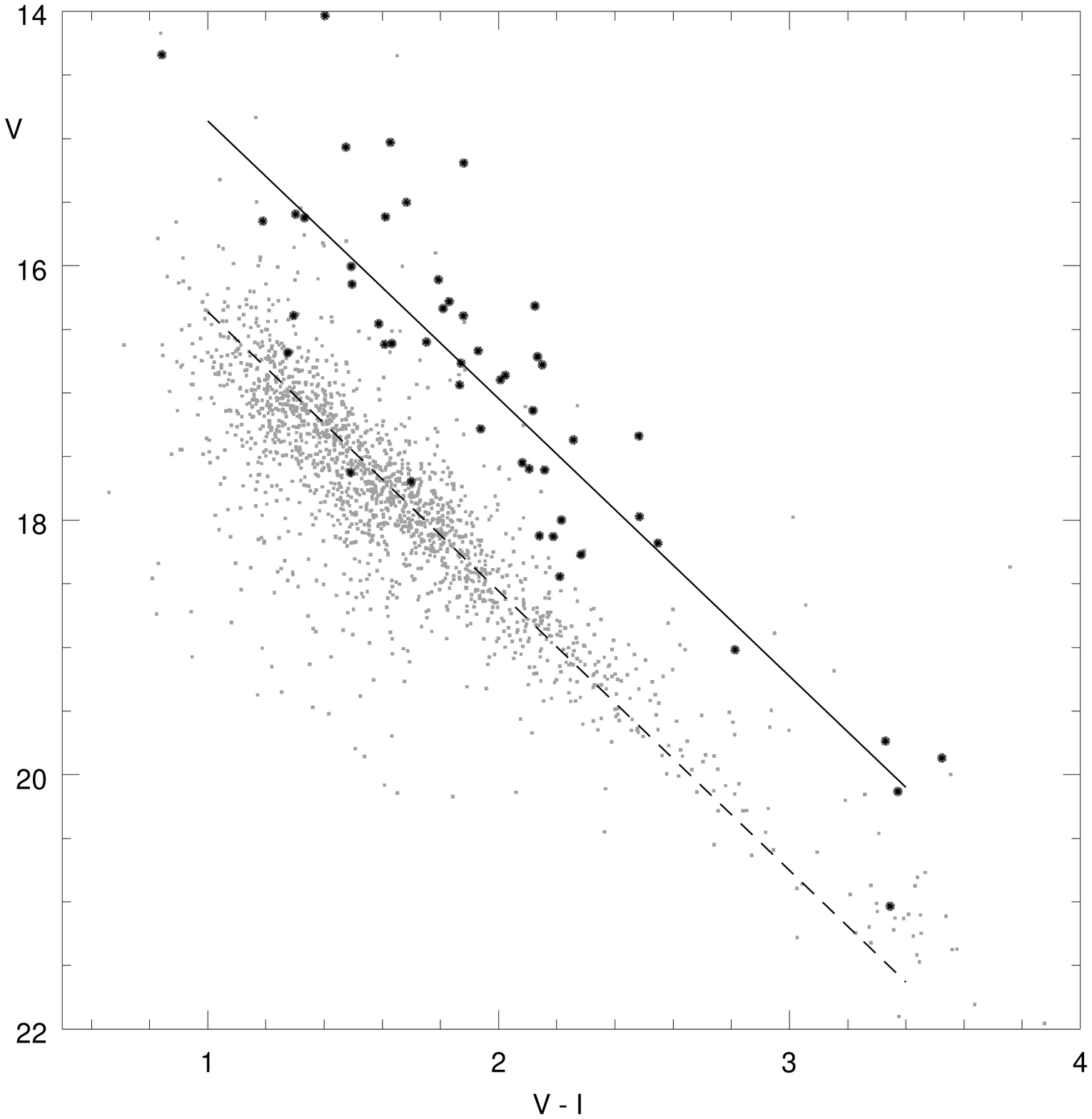}} 
\FigCap{Observed CMD for the Galactic bulge Population~II Cepheids (big dots) 
and RR~Lyr variable stars (grey dots). Solid line is a least square fit to 
the big dots and the broken line is the same line shifted by 1.5~mag in 
brightness $V$.} 
\end{figure} 

It may be also interesting to note the surprisingly close similarity of the 
light curves from Fig.~1 with the light curves of the classical Cepheids 
observed in the Magellanic Clouds. In Fig.~2 we show the normalized 
light curves of classical Cepheids from the Small Magellanic Cloud, with 
periods in the range between about 5 and 30~days, taken from Udalski \etal 
(1999c). In particular, one may note that the Hertzsprung progression, clearly 
seen in Fig.~2 in the period range between about 6 and 24~days, is 
reproduced in Fig.~1 between 0.9 and 3~days. This similarity of light curves 
of classical Cepheids and Population~II Cepheids was first suggested from a 
much poorer observational material by Stobie (1973). 

For further comparison we list in Table~2 the pulsating objects from the LMC 
located below the relation for the fundamental mode Population~I Cepheids 
(Fig.~2 in Udalski \etal 1999b) in the $P$--$L$ diagram and interpreted as 
Population~II Cepheids. Their normalized light curves are shown in Fig~3. 
It should be added that for compatibility with the Galactic bulge objects
their mean photometry was recalculated to be also averaged over magnitudes
instead of intensities as in the original catalog (Udalski \etal 1999b).

The similarity of the shown light curve shapes can be also demonstrated in the 
Fourier parameter diagrams. Fourier analysis appeared to be a convenient way 
of quantitative description of the light curves for classical and short period 
Type~II Cepheids (\eg Simon and Moffett 1985, Petersen and Diethelm 1986). 
Efforts were also undertaken to find convenient criteria of discrimination 
between different types of variable stars in the space of the Fourier 
parameters of their light curves (see \eg Pojma{\'n}ski 2002). Following this 
line we approximated light variations $I(t)$, with periods ${P=2\pi/\omega}$ 
known from periodogram analysis, by a sum of four harmonics: 
$$I(t)=a_0+\sum^4_1a_i\cos(i\cdot\omega t+f_i)$$
Among possible combinations of $a_i$ and $f_i$ parameters, the most suitable 
for separation of objects with different shapes of light curves appeared to 
be: ${r_{21}{=}a_2/a_1}$, ${r_{31}=a_3/a_1}$, ${f_{21}=f_1-2f_1}$ and 
${f_{31}=f_1-3f_3}$. They are shown in Fig.~4 as functions of $\log P$. It can 
be seen that in all panels the distributions of squares, representing the 
classical Cepheids from the SMC shown in Fig.~2, are repeated at 
shorter periods by type~II Cepheids from the Galactic Bulge (dots) and LMC 
(crosses). 

\Section{Period--Luminosity Relation}
The newly identified Population II Cepheids are distributed in many fields 
covered by OGLE-II observations although the vast majority of objects is 
located in the central parts of the Galactic bulge (${-2\arcd<l<2\arcd}$). 
In different fields they are subject to different, sometimes quite large, 
interstellar extinction and we may expect that their apparent color and 
brightness are determined (apart from the shifts resulting from slightly 
different distances and differences in absolute magnitude that can be larger 
as the periods of the objects cover non-negligible range) mainly by the amount 
of interstellar absorption. This is clearly demonstrated by the CMD shown in 
Fig.~5. The solid line in this figure (the best fit to the position of 
Cepheids) is practically the reddening line with equation: 
$$V=12.69(\pm0.36)+2.18(\pm0.17)\cdot(V-I)$$ 
confirming the anomalous value of the ratio ${A_{V}/E(V-I)\approx2}$ in the 
direction of the Galactic bulge as found by Udalski (2003). 

This conclusion is supported by the distribution of RR~Lyr stars in the same 
diagram. Grey dots, representing in Fig.~5 the RR Lyr stars identified by us 
in the Wo{\'z}niak \etal (2002) catalog, follow practically the same reddening 
line as Population~II Cepheids. The broken line in this figure has the same 
slope as the solid line and is shifted toward fainter magnitudes by 1.5~mag. 
The reasonable fit of the shifted line to the grey dots
confirms our conclusion that the CMD for the 
bulge type~II Cepheids is determined in the large part by the interstellar 
extinction. 

On the other hand relatively large scatter of Population~II Cepheids around 
the fitted line (standard deviation of magnitude residuals as large as 
0.68~mag) compared to RR~Lyr stars clearly indicates that the intrinsic 
brightness of the presented sample of Cepheids must significantly vary from 
object to object. This is expected as the range of pulsation periods of the 
stars covers more than order of magnitude. 

To verify how the intrinsic brightness may affect the relation presented in 
Fig.~5 we corrected the {\it VI} magnitudes for the assumed $P$--$L$ relation. 
We tried, as the most likely case, $P$--$L$ relations for the {\it VI} bands 
for Population~II Cepheids from the LMC (Table~2; corrections ${\Delta 
V{=}1.46\log P}$ and ${\Delta I{=}1.85\log P}$). The resulting slope of the fitted 
line to the corrected (${V,V-I}$) positions of Cepheids turned out to be 
${1.97\pm0.08}$ in this case with dramatically smaller scatter (standard 
deviation of magnitude residuals: 0.34~mag). For comparison, we also tried 
other relations, namely $P$--$L$ relations for Classical Cepheids (Udalski 
\etal 1999a; corrections ${\Delta V=2.76\log P}$ and ${\Delta I=2.96\log P}$). 
The resulting slope was similar -- within 0.15 of the value obtained for the 
LMC Population~II Cepheids, but the standard deviation of residuals was again 
much larger (0.50~mag). 

All our tests consistently show that the position of our sample of 
Population~II Cepheids in the CMD diagram indicates non-standard interstellar 
extinction in the Galactic bulge. The direction of the reddening line is fully 
consistent with the direction found with completely different type of stars. 
The slope of the reddening line in the OGLE-II filters resulting from our 
Cepheid sample is in the range of 1.9--2.1 in ideal agreement with red clump 
stars (Udalski 2003). Direction of the reddening line is also in very good 
agreement with the direction indicated by RR~Lyr stars. The linear fit with 
$2.5\sigma$ clipping algorithm to remove outliers to grey points in Fig.~5 
representing location of RR Lyr stars in the CMD diagram yields the slope of 
1.9 with standard deviation of magnitude residuals equal to 0.30~mag. 

Once the character of the interstellar extinction toward the Galactic bulge is 
known we may calculate the extinction free index: 
$$W_{V}=V-R(V-I)$$
and determine $P$--$L$ relation for our sample of Population~II Cepheids. It 
should be stressed that due to large and variable interstellar extinction, 
determination of the $P$--$L$ relation for individual bands would be of very 
poor quality in the Galactic bulge case. 

To find the best $P$--$L$ relation for the $W_{V}$ index we performed 
additional test. We varied the $R$ coefficient within the range
determined  above to obtain the lowest scatter of the ${\log P-W_V}$
relation. One has to be careful with such a procedure when the
interstellar reddening is comparable with the width of the instability
strip because simultaneously a scatter along the line of constant period
is minimized. As a result the determined $R$ does not directly
correspond to the real coefficient of interstellar reddening. In the
case of the Galactic bulge, however, the interstellar reddening is
always by a factor of several larger than the width of the instability
strip, therefore this effect is minimal.  The  standard deviation of
residuals turned out to be the lowest in our case for ${R=2.0}$. 

\begin{figure}
\vglue-7mm
\centerline{\includegraphics[width=6cm]{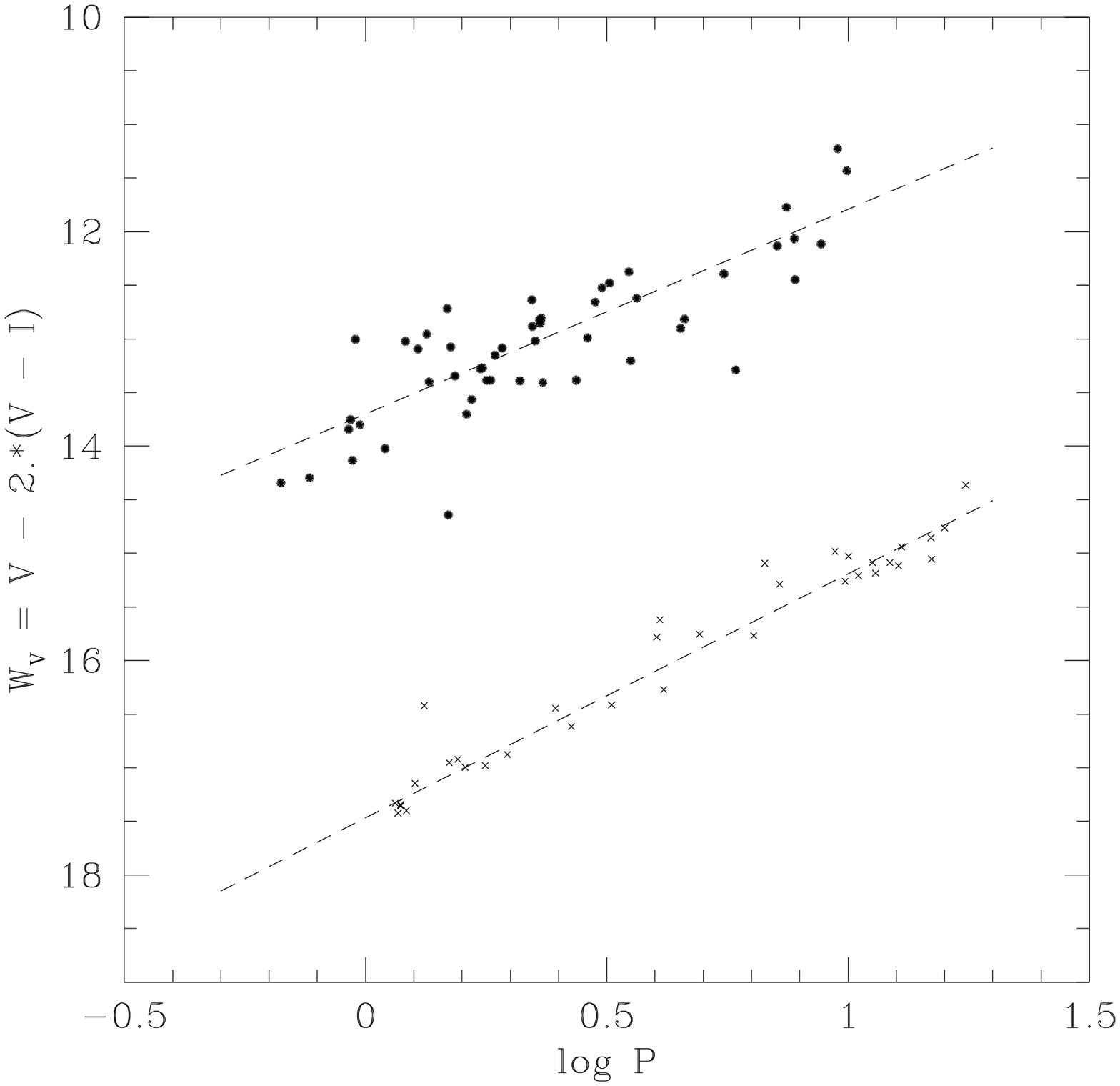}} 
\FigCap{$W_V$ index ($R=2.0$) as a function of $\log P$ for the Galactic bulge 
Population~II Cepheids (big dots) and Population~II Cepheids from the Large 
Magellanic Cloud (crosses). Broken lines represent the best linear least 
square fit to the points, \ie the $P$--$L$ relations.} 
\end{figure} 
The dependence of the $W_V$ index on $\log P$ for Population~II Cepheids in 
the Galactic bulge is shown in Fig.~6. The solid line in this figure is the 
linear least square fit to the dark dots with $2.5\sigma$ clipping algorithm 
to remove outliers: 
$$W_V=13.70(\pm 0.07)-1.91(\pm 0.16)\cdot\log P$$
with the standard deviation of residuals equal to 0.32~mag and the number
of Cepheids: ${n=49}$ (two outliers removed by the clipping algorithm).

It is interesting to compare our relation with the similar one for the
LMC. In  the case of the LMC the interstellar extinction seems to follow
the standard  relation with ${R\approx2.5}$ (Udalski 2003). Therefore
the $W_V$ index  calculated with ${R=2.0}$ is not extinction
independent. Nevertheless the  reddening in the LMC is relatively small
and the {\it VI} photometry for  Population~II Cepheids could be
reasonably dereddened (Udalski \etal 1999ab).  Thus in the case of the
LMC the $W_V$ index (for ${R=2.0}$) was calculated from the dereddened 
values $V_0$ and $I_0$, presented in Table~2. 

The relation for the LMC type II Cepheids is shown by crosses in Fig.~6. The 
linear least square fit to the crosses, again with the $2.5\sigma$ clipping 
algorithm, is: 
$$W_V=17.47(\pm0.05)-2.28(\pm0.06)\cdot\log P$$
with the standard deviation of only 0.14~mag. The scatter of points in the 
case of the bulge Population~II Cepheids is markedly bigger than in the case 
of the LMC and this can be understood as a consequence of possible 
irregularities in extinction toward the Galactic bulge as also of some 
differences in distance to particular objects. 

The two slope values differ somewhat but the difference is only at $2.1\sigma$ 
level. The values of slopes of both $P$--$L$ relations could converge if we 
lowered the coefficient $R$ in the definition of $W_V$ to 1.4. However, the 
scatter of both relations would be then unacceptably large: by a factor of two 
bigger than for ${R=2.0}$. 

We may conclude that at this stage the difference of the slopes of
$P$--$L$  relation for Population~II Cepheids is not significant,
although larger samples of objects from both stars systems are necessary
for final confirmation. If the zero point of  the $P$--$L$ relation is
also constant then Population~II Cepheids could be  used as standard
candles for distance determinations similarly to Classical  Cepheids. 

\vspace*{6pt}
\Section{Discussion}
\vspace*{3pt}
The main results presented in this paper can be summarized as follows:
\begin{enumerate}
\item The Galactic bulge contains Population II Cepheids.
\item Type II Cepheids are on average by about 1.5~mag brighter than the 
average brightness of the Galactic bulge RR~Lyr stars. 
\item The shapes of light curves of short period Population~II Cepheids are, 
in general, very similar to the shapes of classical Cepheids pulsating with 
much longer periods. 
\item The shapes of Population~II Cepheids light curves differ among 
themselves in the same way as differ also light curves of classical Cepheids: 
most probably in both cases the differences of light curve shapes of stars 
with the same period are a consequence of somewhat different chemical 
composition. 
\item Population~II Cepheids in the Galactic bulge are significantly reddened 
by interstellar extinction. Their position in the CMD diagram confirms 
anomalous interstellar extinction toward the Galactic bulge (Udalski 2003) in 
very good agreement with the results obtained with red clump and RR~Lyr stars. 
\item All Population II Cepheids follow the same $P$--$L$ relation and in this 
sense they form a physically homogeneous group what confirms the McNamara 
(1995) result concerning the objects with periods shorter than 10~days. We did 
not find in the Galactic bulge the Population~II Cepheids with periods longer 
than 10~days. Such objects, however, are observed in the LMC (Fig.~6). 
Contrary to McNamara (1995) finding they follow the same $P$--$L$ relation as 
the Population~II Cepheids with shorter periods. 
\item The comparison of slopes of the $W_V$ index $P$--$L$ relation
for  the Galactic bulge and LMC indicates that both are within
the errors  of determination the same. Thus, if the zero point of the
$P$--$L$ relation is  constant the type~II Cepheids might be used as
standard candles. It is  tempting to make an exercise of determination 
on the base of our results the difference of distance  moduli between
the Galactic bulge and LMC. The average difference of the $W_V$  index
for the Galactic bulge and LMC Cepheids is about ${3.6\pm0.2}$~mag what 
with the geometric distance modulus to the Galactic bulge,
${14.5\pm0.1}$~mag,  (Eisenhauer \etal 2003) leads to the short distance
modulus to the LMC.  However, one has to remember that such a
determination might be significantly  biased. The change of the $R$
coefficient in the $W_V$ definition by only 0.1  moves the zero point of
the Galactic bulge relation by as much as 0.2~mag and  only by a few
hundredths of magnitude for the LMC relation due to large difference of ${V-I}$
colors (the slopes change marginally). Therefore, to perform reliable 
comparison the coefficient $R$ should be determined with high precision.
Also  other systematic effects cannot be excluded. 
\item Relatively large sample of Population~II Cepheids in the Galactic bulge 
makes it possible to study their distribution within the bulge. For instance, 
the red clump stars are located in the Galactic bar what shows up as almost 
0.4~mag difference in the mean brightness of red clump stars in the fields located 
at ${l=\pm5\arcd}$ (Stanek \etal 1994). On the other hand RR~Lyr stars in the 
Galactic bulge do not show such asymetry and are located in the inner 
extension of the Galactic halo (Alcock \etal 1998). 

\begin{figure}
\vglue-7mm
\centerline{\includegraphics[bb=20 35 519 415, width=7cm]{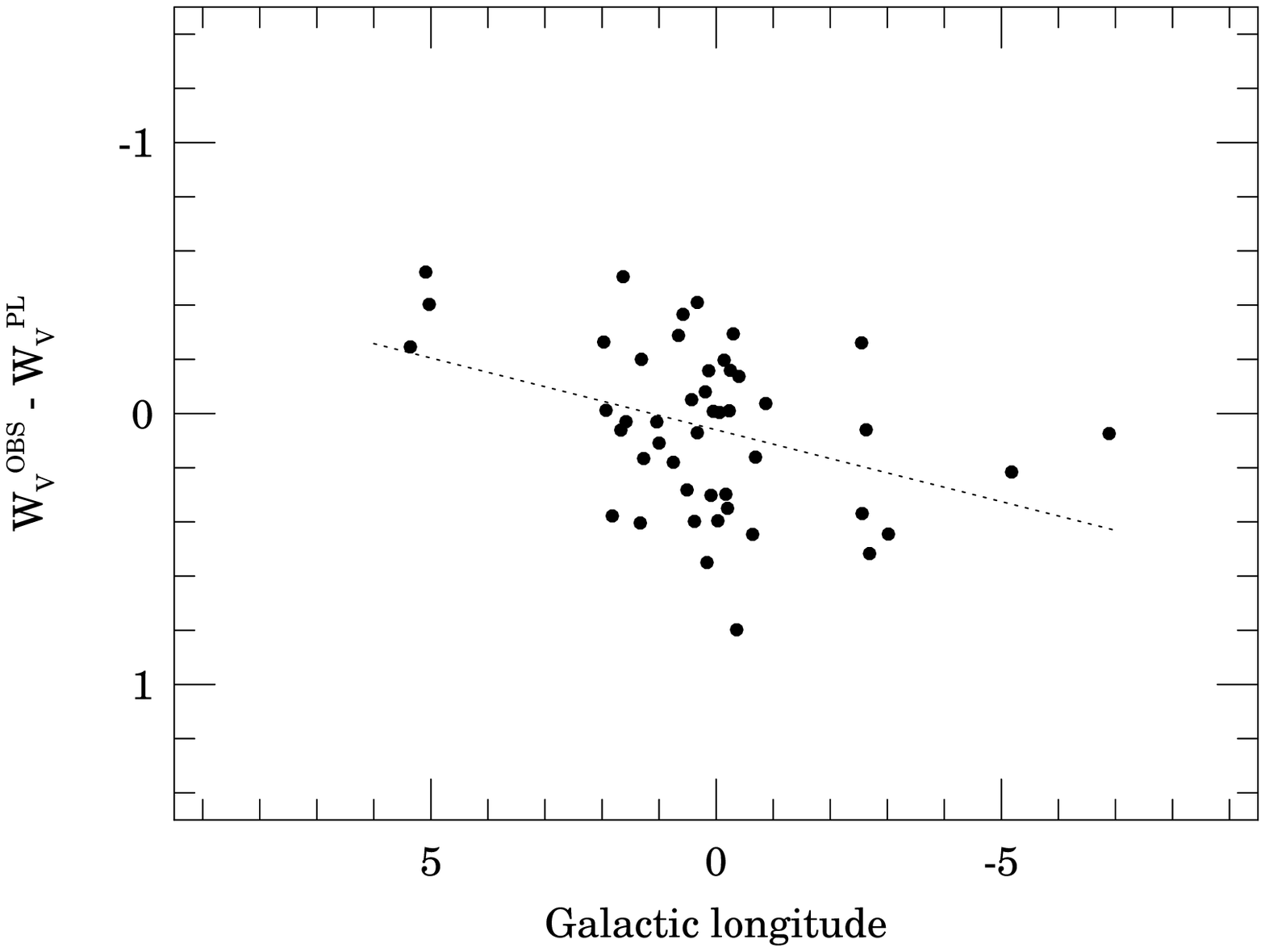}} 
\FigCap{Deviation of the observed $W_V$ index from the $P$--$L$ relation as a 
function of the Galactic longitude. Dotted line shows the best linear fit to 
the data.}
\end{figure} 
To study the distribution of Population~II Cepheids in the Galactic bulge we 
plotted the deviations of the $W_V$ index from the mean $P$--$L$ relation as a 
function of the Galactic longitude (Fig.~7). Although the vast majority of 
objects are located relatively close to ${l\approx0\arcd}$ there is a visible 
trend indicating that the objects with positive $l$ are on average somewhat 
brighter than the mean $P$--$L$ relation while those with negative $l$ are 
fainter. The trend is marked with a dotted line in Fig.~7 (the best linear
fit) and reaches about 0.5~mag at ${l=\pm5\arcd}$. 

The observed effect is identical as observed for the red clump stars. Also its 
magnitude is very similar. Thus it seems natural to interpret it similarly 
to red clumps stars case, \ie that the Population~II Cepheids are distributed 
in the Galactic bar. However, one has to remember that the number of objects 
with ${l>2\arcd}$ and ${l<-2\arcd}$ is small in the present sample. Therefore 
additional detections of Population~II Cepheids are necessary for full 
confirmation of our finding. 
\end{enumerate}

It is worth noting that the presented in this paper Population~II Cepheids 
come from the second phase of the OGLE project. Currently the OGLE survey 
regularly monitors by almost an order of magnitude larger area in the Galactic 
bulge and Magellanic Clouds. Therefore, one can expect that the number of 
detected Population~II Cepheids in these star systems should increase significantly 
in the near future and presented in this paper relations will be verified with 
much larger samples. 

\Acknow{We would like to thank Drs.\ W.\ Dziembowski and P.\ Moskalik for 
their helpful and clarifying remarks on the subjects of this paper. We also 
thank Prof.\ B.\ Paczy{\'n}ski for many interesting comments. 

This work was supported by the KBN grant BST to the Warsaw University 
Observatory and by the KBN grant 2P03D02124 to A.\ Udalski. Partial support to 
the OGLE project was provided with the NSF grant AST-0204908 and NASA grant 
NAG5-12212 to B.\ Paczy\'nski.}

\end{document}